\documentclass[
 aip,        
 jcp,        
 amsmath,amssymb,
 reprint,    
]{revtex4-1}

\usepackage{graphicx}      
\usepackage{bm}            
\usepackage[utf8]{inputenc}
\usepackage[T1]{fontenc}
\usepackage{mathptmx}
\usepackage{etoolbox}
\usepackage{algorithm}
\usepackage{appendix}
\usepackage{xcolor}
\usepackage{algpseudocode}  
\usepackage{array}[=2016-10-06]
\usepackage{dcolumn}

\makeatletter
\def\@email#1#2{%
 \endgroup
 \patchcmd{\titleblock@produce}
  {\frontmatter@RRAPformat}
  {\frontmatter@RRAPformat{\produce@RRAP{*#1\href{mailto:#2}{#2}}}\frontmatter@RRAPformat}
  {}{}
}%
\makeatother

\begin{document}


\title{A differentiable photon-by-photon likelihood for continuous free-energy landscapes and diffusion coefficients from single-molecule FRET}

\author{Lars Dingeldein}
\affiliation{Institute of Physics, Goethe University Frankfurt, Frankfurt am Main, Germany}
\affiliation{Frankfurt Institute for Advanced Studies, Frankfurt am Main, Germany}

\author{Roberto Covino}
\email{covino@fias.uni-frankfurt.de}
\affiliation{Frankfurt Institute for Advanced Studies, Frankfurt am Main, Germany}
\affiliation{Institute of Computer Science, Goethe University Frankfurt, Frankfurt am Main, Germany}
\affiliation{Cluster of Excellence SubCellular Architecture of Life (SCALE), Goethe University Frankfurt, Frankfurt, Germany}

\date{\today}

\begin{abstract}
Single-molecule FRET probes the conformational dynamics of biomolecules by measuring the distance between two dyes. The experiment produces a stream of coloured photons, which is only an indirect readout of these dynamics. Recovering the free-energy landscape and diffusion coefficient from such a photon stream is a difficult inverse problem. Existing approaches assume a small number of discrete states, bin the photons, or are computationally expensive. Here we derive an exact likelihood for the recorded photon stream under a model in which the dye distance diffuses on a continuous free-energy landscape. It uses the raw inter-photon times and colours at full time resolution, and analytically integrates out all hidden trajectories. The likelihood is differentiable, so automatic differentiation returns exact gradients with respect to all model parameters. This lets us jointly infer the free-energy landscape, the diffusion coefficient, and the photophysical parameters by gradient-based optimization. On simulated data, we recover free-energy landscapes, including ones with a short-lived intermediate, together with the diffusion coefficients. Uncertainties follow from the curvature of the likelihood, computed directly from the same gradients. The framework also guides experimental design. Before any data are recorded, we can evaluate how much a given acquisition setting reduces the resulting uncertainty. Evaluation on the GPU is fast, and independent traces are processed in parallel, so a single fit converges in minutes and scales to large datasets. The likelihood extends photon-by-photon analysis to continuous free-energy landscapes and diffusion coefficients, putting fast quantitative inference with uncertainties within reach of the smFRET community.
\end{abstract}

\maketitle

\section{Introduction}
\label{sec:introduction}
\begin{figure*}
\centering
\includegraphics[width=\textwidth]{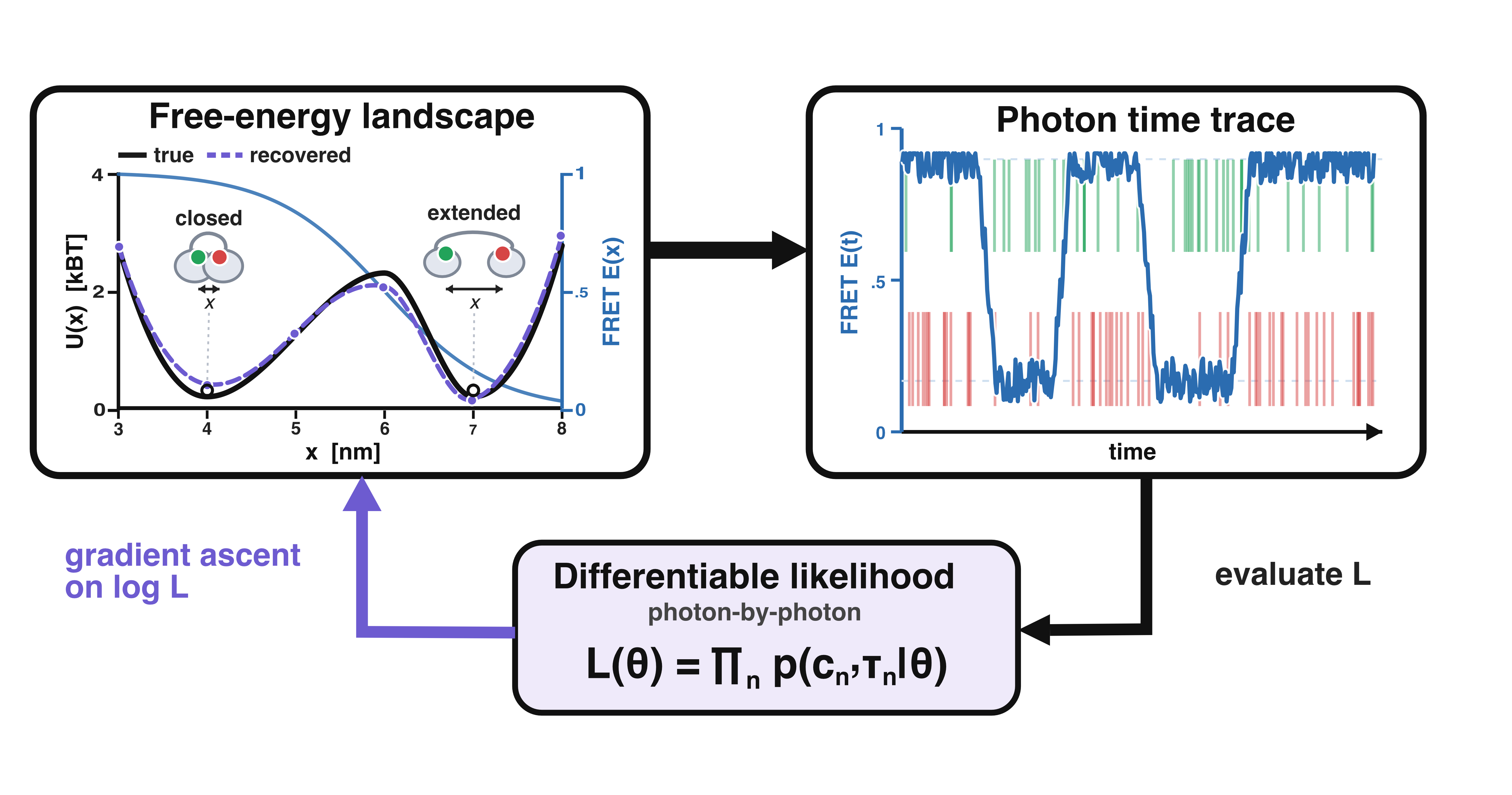}
\caption{\label{fig:fig1} \textbf{Photon-by-photon inference of a diffusion model.} A molecule diffuses on a one-dimensional free-energy surface $u(x)$ (left, black), where the donor-acceptor distance $x$ sets the FRET efficiency $E(x)$ (blue). The resulting dynamics emit a stream of donor/acceptor photons with channels $c_n$ and inter-photon times $\tau_n$ (right). A candidate model $\boldsymbol{\theta}=\{u(x),D,\text{photophysics}\}$ is scored by the differentiable photon-by-photon likelihood $L(\boldsymbol{\theta})=\prod_n p(c_n,\tau_n\mid\boldsymbol{\theta})$ (bottom). Automatic differentiation yields exact gradients, and gradient optimization on $\log L$ updates the model until the recovered landscape (left, dashed) matches the truth.}
\end{figure*}

The conformational dynamics of biomolecules are integral to their biological function, so characterizing the metastable states a molecule visits, their relative energetics, and the timescales that connect them is central to understanding how biomolecules work. Single-molecule FRET (smFRET) is a powerful method to observe such dynamics directly on individual biomolecules\cite{smfret_review, schuler2016single}. The conformational dynamics are probed by measuring the distance between two fluorescence dyes that are attached to a biomolecule. A laser excites the so-called donor dye, which transfers energy to the acceptor dye through a distance-dependent dipole coupling\cite{forster1948zwischenmolekulare}. Because both dyes emit energy in the form of different colored photons --- green for the donor, red for the acceptor --- their relative frequency reports on the distance between them. The F\"orster law relates the transfer efficiency to this distance \cite{forster1948zwischenmolekulare},
\begin{equation}\label{eq:forster}
  E(x) = \frac{R_0^6}{R_0^6 + x^6},
\end{equation}
where $x$ is the donor-acceptor distance and $R_0$ is the F\"orster radius. When the dyes are close, energy transfer from donor to acceptor is efficient, and the photons are mostly red. When the dyes are far apart, transfer is weak, and the photons are mostly green. 
Since we probe conformational dynamics by the donor-acceptor distance $x(t)$, which changes over time, we observe a stream of colored photons where the relative frequency of green and red photons is changing (Fig. \ref{fig:fig1},  photon time trace). The photon stream is therefore only an indirect and noisy readout of the changing donor-acceptor distance $x(t)$ (Fig.~\ref{fig:fig1}).

For discrete-state kinetics Gopich and Szabo formulated a photon-by-photon maximum likelihood approach ~\cite{gopich_szabo}. Continuous landscapes have since been recovered by Bayesian nonparametrics~\cite{skipperfret, Sgouralis2019} and expectation--maximization~\cite{em_smfret}. These approaches, however, either bin photons, sample the hidden trajectory explicitly or can be computationally expensive.

Here, we follow the approach pioneered by Gerhard Hummer \cite{hummer2005position}, and build on a maximum likelihood fit of a dynamical diffusive model to data. 
We model the dynamics of the donor-acceptor distance $x(t)$ by Brownian diffusion on a free-energy landscape $u(x)$ with a diffusion coefficient $D$ \cite{best2005reaction, best2010coordinate, hummer2005position}. Learning such a minimal model of the dynamics underlying $x(t)$ from the raw photon stream then reduces to fitting the Brownian dynamics model to the photon stream (Fig. \ref{fig:fig1}, free energy landscape)
First, we derive an exact photon-by-photon likelihood that is end-to-end differentiable. R0epresenting $u(x)$ as a smooth parametric function, we estimate the free energy landscape, $D$, and the photophysics parameters jointly by gradient based optimization using automatic differentiation (Fig. \ref{fig:fig1}, differentiable likelihood). On simulated data, the method recovers ground-truth landscapes, including short-lived intermediates. It scales to $\sim\!10^6$ photons on a single GPU. Since the likelihood and its gradients are cheap to evaluate, we  can efficiently quantify the uncertainties.

\section{Theory}
\label{sec:theory}

In a single-molecule FRET experiment, we measure the arrival times and colors of individual photons,
$\mathcal{D}=\{(t_n,c_n)\}_{n=1}^{N}$ with $0<t_1<\dots<t_N<T$ and $c_n\in\{g,r\}$. Here, $t_n$ is the arrival time of the $n$th photon and $c_n$ is its color, with $g$ the green donor channel and $r$ the red acceptor channel. The $T$ long trace contains $N$ photons in total. For several traces we write $\mathcal{D}_m$, $m=1,\dots,N_{\mathrm{trace}}$, for the $m$th record. Throughout, we consider continuous illumination of a single molecule at thermal equilibrium.

 We assume that $\mathcal{D}$ was produced by the hidden donor-acceptor distance $x(t)$, which can be well described by one-dimensional Brownian diffusion\cite{hummer2005position}. Our goal is to infer the free-energy landscape $u(x)$ and the diffusion coefficient $D$ that best fit $\mathcal{D}$.  To do that, we need to maximize the likelihood $L(\boldsymbol{\theta}) = p(\mathcal{D};\boldsymbol{\theta})$ of observing $\mathcal{D}$ given the model parameters $\boldsymbol{\theta}  = \{u(x), D, \mathrm{photophysics}\}$.
 
\subsection{Forward model}
\label{subsec:forward_model}

The dynamics of the hidden donor-acceptor distance $x(t)$  can be described by the overdamped Langevin equation,
\begin{equation}\label{eq:langevin}
dx = -D\,u'(x)\,dt + \sqrt{2D}\,dW_t .
\end{equation}
Here, $u(x)=U(x)/k_{\mathrm{B}}T$ is the free-energy landscape made dimensionless by setting $k_{\mathrm{B}}T=1$. The diffusion coefficient $D=k_{\mathrm{B}}T/\zeta$ is fixed by the friction coefficient $\zeta$ through the Einstein relation. Finally, $W_t$ is a standard Wiener process, so that $\sqrt{2D}\,dW_t$ is a Gaussian step of variance $2D\,dt$. 

To track the probability density of finding the coordinate at $x$ at time $t$, we use the Fokker--Planck
equation $\partial_t p=\mathcal{L}p$, with the generator
\begin{equation}\label{eq:generator}
  \mathcal{L}p = D\,\partial_x\!\big[e^{-u}\,\partial_x(e^{u}p)\big]
  = D\,\partial_x\big(\partial_x p + p\,u'\big) .
\end{equation}
Here, $\mathcal{L}$ is the forward (Smoluchowski) generator. The stationary solution $\partial_t p=0$ with reflecting boundaries is the Boltzmann density $\pi_{\boldsymbol{\theta}}(x)\propto e^{-u(x)}$. We assume the molecule starts from the equilibrium distribution $x(0)\sim\pi_{\boldsymbol{\theta}}$.
 
The Förster law of Eq.~\eqref{eq:forster} converts the distance $x$ into transfer efficiency,  but the detector registers photon arrival events and not efficiencies. Additionally, real optics are imperfect, and the photons reach us filtered through detection efficiencies, spectral crosstalk, and background \cite{Saurabh2023}. Each channel collects a 
mixture of donor and acceptor emission.
\begin{equation}
\label{eq:photon_rates}
\begin{aligned}
  \lambda_g(x) &= a_{g\to g}\,[1-E(x)] + a_{r\to g}\,E(x) + \beta_g ,\\
  \lambda_r(x) &= a_{g\to r}\,[1-E(x)] + a_{r\to r}\,E(x) + \beta_r .
\end{aligned}
\end{equation}
Here, $\lambda_c(x)$ is the rate, at which we detect a photon in channel $c$ while the  donor-acceptor distance is at $x$. The total detection rate is $\lambda_{\mathrm{tot}}(x)=\lambda_g(x)+\lambda_r(x)$.  Each amplitude $a_{c'\to c}=\eta_c C_{c'\to c} k_{\mathrm{em}}$ includes three instrumental factors: the donor emission rate $k_{\mathrm{em}}$, the detection efficiency $\eta_c$ of channel $c$, and the crosstalk element $C_{c'\to c}$ for a photon emitted by dye $c'$ that lands in channel $c$. We fix the Förster radius $R_0$ and the crosstalk elements $C_{c'\to c}$ throughout, since both are calibrated beforehand. The constants $\beta_g$ and $\beta_r$ are the background rates from scatter and dark counts, which do not depend on the molecule. Only these six numbers enter the likelihood: four amplitudes and two backgrounds, from which only the two products $a_c = \eta_c\cdot k_{\mathrm{em}}$ and the two background parameters $\beta_c$ are optimized. 

\subsection{Marginalizing the hidden trajectory}
\label{subsec:marginalization}

For a fixed trajectory, the photon record is an inhomogeneous Poisson
process whose rates are read out along the path~\cite{DaleyVereJones2003}. Its probability
density---the path-conditional likelihood---is
\begin{equation}
  p\!\left(\mathcal{D}; x(\cdot),\boldsymbol{\theta}\right)
  = \prod_{n=1}^{N}\lambda_{c_n}\!\big(x(t_n)\big)\,
    e^{-\int_0^T \lambda_{\mathrm{tot}}(x(t))\,dt}.
  \label{eq:cond-lik}
\end{equation}
We never observe $x(t)$. The likelihood follows by averaging
Eq.~\eqref{eq:cond-lik} over every trajectory the molecule could have
taken, each weighted by its probability under the path density
$\mathbb{P}_{\boldsymbol{\theta}}$ of Eq.~(2), started from
equilibrium, $x(0)\sim\pi_{\boldsymbol{\theta}}$:
\begin{equation}
  L(\boldsymbol{\theta})\equiv p\!\left(\mathcal{D};\boldsymbol{\theta}\right)
  = \mathbb{E}_{x(\cdot)\sim\mathbb{P}_{\boldsymbol{\theta}}}
    \!\left[\,p\!\left(\mathcal{D}; x(\cdot),\boldsymbol{\theta}\right)\right].
  \label{eq:lik}
\end{equation}
The path density depends on $\boldsymbol{\theta}$ only through the landscape $u$ and the diffusion coefficient $D$ and the photon rates only through the detection coefficients $a_{c'\to c}$ and backgrounds
$\beta_c$. 

Evaluating Eq. \ref{eq:lik} by explicitly running many trajectories $x(t)$ and averaging is computationally expensive. However, here we show that this is not necessary, since we can actually solve it recursively. The main idea is to propagate a single density through the observed photon trace. The density propagates through each dark interval while being reweighted at each detected photon.. This is enough because $x(t)$ is Markovian. The future position of each path depends only on the current position, so we never need the history. We therefore never need to know the path's history, only the position at the photon detection, the density at the last photon. We call this density the forward filter $\rho(x,t)$: the quantity $\rho(x,t)\,dx$ is the joint probability that the coordinate lies in $[x,x+dx]$ at time $t$ and that the recorded photons occurred in $[0,t]$ so far. We can write the forward filter density as a path integral, where we weigh each path by the conditional likelihood
\begin{equation*}
  \rho(x,t)=\mathbb{E}_{x(\cdot)\sim\mathbb{P}_{\boldsymbol{\theta}}}\!\big[\,
  \delta\big(x(t)-x\big)\,\Gamma_t\,\big] ,
\end{equation*}
where $\delta$ is the Dirac delta and $\Gamma_t$ is the conditional weight of Eq.~\eqref{eq:cond-lik} accumulated up to time $t$,
\begin{equation*}
  \Gamma_t=\bigg[\prod_{n:\,t_n\le t}\lambda_{c_n}\big(x(t_n)\big)\bigg]\,
  e^{-\int_0^t\lambda_{\mathrm{tot}}(x(s))\,ds}  .
\end{equation*}
In other words, the delta selects all the paths that end up in $x$, while $\Gamma_t$ weights them by how probable the already observed photons are. Running $\rho$ from $t=0$ to $t=T$ evaluates the entire trace and integrating out the unobserved final position turns the sub-density into the final likelihood.
\begin{equation}\label{eq:filter}
  L(\boldsymbol{\theta})=\int\rho(x,T)\,\mathrm{d}x = \langle\mathbf{1},\rho(\cdot, T)\rangle,
\end{equation}
with $\rho(\cdot,0)=\pi_{\boldsymbol{\theta}}$, where we introduced the inner-product notation because it will be convenient later on.
 
Computing the likelihood can now be separated into sequential operations along the trace. First, consider a dark interval. In a small step $dt$, two things happen to $\rho$. The coordinate diffuses, so $\rho\to\rho+\mathcal{L}\rho\,dt$. And no photon is fired, which multiplies $\rho$ by the local survival probability $1-\lambda_{\mathrm{tot}}(x)\,dt$. Composing the two terms gives a killed Fokker--Planck equation, together with its propagator,
\begin{equation}\label{eq:killedFP}
  \begin{aligned}
    \partial_t\rho &= (\mathcal{L}-\lambda_{\mathrm{tot}})\,\rho , \\
    \mathcal{S}(\tau) &\equiv e^{(\mathcal{L}-\lambda_{\mathrm{tot}})\tau} .
  \end{aligned}
\end{equation}
Here, $\lambda_{\mathrm{tot}}(x)$ acts as a position-dependent killing rate of the density. Probability is destroyed wherever a photon would likely have fired, which is precisely the statement that the interval was dark. Formally, $\mathcal{S}$ is the Feynman-Kac propagator for a diffusion killed at rate $\lambda_{\mathrm{tot}}$~\cite{KaratzasShreve1991}, which solves the killed equation over a gap of duration $\tau$. 

At a detection of a photon with color  $c_n$ we just need to reweigh the forward-filter density by $\lambda_{c_n}(x)$.
\begin{equation}\label{eq:jump}
  \rho(x,t_n^+)=\lambda_{c_n}(x)\,\rho(x,t_n^-)\equiv(\Lambda_{c_n}\rho)(x).
\end{equation}
Here, $t_n^-$ and $t_n^+$ mean just before and just after the photon, and $\Lambda_c$ is the emission operator, which is nothing but multiplication by $\lambda_c(x)$.
 
\subsection{Forward filtering}
\label{subsec:forward_filtering}

 Now we can construct the likelihood over the entire path by chaining two events, propagation of the density during dark intervals, with reweighting at the photons.  Let the inter-photon intervals be $\tau_1=t_1$, $\tau_n=t_n-t_{n-1}$ for $n=2,\dots,N$ and $\tau_{N+1}=T-t_N$, so that the gaps chunk the observation window $\sum_{n=1}^{N+1}\tau_n=T$. We start from the equilibrium density, alternate the dark-gap propagator $\mathcal{S}$ with the photon reweighting $\Lambda_c$, and integrate out the final position. This leads to the exact likelihood,
\begin{equation}\label{eq:oplik}
  \begin{split}
    L(\boldsymbol{\theta}) = \big\langle \mathbf{1},\
        &\mathcal{S}(\tau_{N+1})\,\Lambda_{c_N}\,\mathcal{S}(\tau_N)\cdots \\
        &\Lambda_{c_1}\,\mathcal{S}(\tau_1)\,\pi_{\boldsymbol{\theta}} \big\rangle .
  \end{split}
\end{equation}
The equation can be read from right to left. Start at the equilibrium distribution. Diffuse and survive for $\tau_1$. See the first photon and reweight by its color. Diffuse and survive for $\tau_2$, and so on through the
last photon. Propagate through the trailing dark stretch $\tau_{N+1}$, then integrate over the final position.
The two boundary objects handle the two positions we never observe. On the right, $\pi_{\boldsymbol{\theta}}$ averages over the unknown starting position. On the left, $\mathbf{1}$ sums over the unknown final position.
 
Haas et al. (2013) obtained the same expression using the Feynman-Kac theorem~\cite{em_smfret}. Eq.~\eqref{eq:oplik} reduces to the photon-by-photon likelihood of Gopich and Szabo by coarse graining the continous $x$ in a finite number of discrete states~\cite{gopich_szabo}.

\section{Methods}
\label{sec:methods}

\subsection{Likelihood evaluation and implementation}
\label{subsec:likelihood_eval}
Equation~\eqref{eq:oplik} is the exact likelihood, but the operators act on functions of the continuous coordinate $x$. We therefore discretise $x$ on a uniform grid of $M$ points $\{x_i\}_{i=1}^M$ with spacing $h = x_{i+1} - x_{i}$, turning every object in Eq.~\eqref{eq:oplik} into a matrix or a vector. The forward-filtered density $\rho$ becomes a vector $\alpha$ of cell probabilities, $[\alpha]_i\approx h\,\rho(x_i,t)$, written $\alpha_n$ for the filter after photon $n$, and the pairing $\langle\mathbf{1},\cdot\rangle$ becomes $\mathbf{1}^{\top}\alpha$. The generator $\mathcal{L}$ becomes an $M\times M$ rate matrix $Q$. $\Lambda_c$ and $\lambda_{\mathrm{tot}}$ multiply pointwise, which on a grid means multiplying with a diagonal matrix, $\Lambda_c=\operatorname{diag}(\lambda_c(x_i))$ and $\Lambda_{\mathrm{tot}}=\operatorname{diag}(\lambda_{\mathrm{tot}}(x_i))$, with $\Lambda_g+\Lambda_r=\Lambda_{\mathrm{tot}}$ entry by entry. The killed propagator of Eq.~\eqref{eq:killedFP} becomes $\mathcal{S}(\tau)=e^{(Q-\Lambda_{\mathrm{tot}})\tau}$, so the likelihood is a chain of matrix--vector products on $\alpha$, started from $\alpha_0=\pi_{\boldsymbol{\theta}}$. The initial condition $\pi_\theta$ is discretized in the same way as $\rho$.

We build $Q$ first, since it is the matrix that represents $\mathcal{L}$ on the grid. A finite-difference discretisation of Eq.~\eqref{eq:generator} converges as $h\to0$, but at finite $h$ its stationary state is not exactly $e^{-u}$, so $\alpha_0=\pi_{\boldsymbol{\theta}}$ would not be stationary and the symmetrization we need later for the discretisation of $\mathcal{S}(\tau)$ would fail. We therefore use the square-root approximation~\cite{bicout1998electron, Donati2021}, which discretises the bracketed form of Eq.~\eqref{eq:generator} as a hopping process between neighboring cells. The rate $q_{i\rightarrow i\pm1}$ to transition from cell $i$ to one of its neigbouring cells $i\pm1$ can be written as
\begin{equation}\label{eq:sqra}
  q_{i\to i\pm1}=\frac{D}{h^{2}}\,
  \exp\!\left(-\frac{u_{i\pm1}-u_i}{2}\right) ,
\end{equation}
with $u_i=u(x_i)$. The tridiagonal rate matrix $Q$ collects these rates, $Q_{i\pm1,i}=q_{i\to i\pm1}$ and $Q_{ii}=-(q_{i\to i-1}+q_{i\to i+1})$, with reflecting boundaries. Two properties hold exactly on the grid, not only as $h\to0$. First the columns of $Q$ sum to zero, so probability is conserved. Second, detailed balance holds with $\pi_{\boldsymbol{\theta},i}\propto e^{-u_i}$, so the Boltzmann distribution is the exact stationary state of $Q$ on the grid. As $h\to0$, $Q$ converges to $\mathcal{L}$.

Now that we have $Q$, we can construct a propagator on the grid. Equation~\eqref{eq:oplik} needs $\mathcal{S}(\tau)=e^{(Q-\Lambda_{\mathrm{tot}})\tau}$  for each of the $N-1$ inter-photon gaps, and forming a matrix exponential that many times is prohibitive. We instead diagonalize $Q-\Lambda_{\mathrm{tot}}$ once. With $\Pi =\operatorname{diag}(\pi_{\boldsymbol{\theta}})$, we can use the similarity transform
\begin{equation}\label{eq:symm}
  A=\Pi^{-1/2}(Q-\Lambda_{\mathrm{tot}})\,\Pi^{1/2}=\Psi\operatorname{diag}(\nu)\,\Psi^{\top}
\end{equation}
that turns the non-symmetric $Q-\Lambda_{\mathrm{tot}}$ into a symmetric tridiagonal matrix $A$~\cite{Prinz2011}, with its orthonormal eigenvectors $\Psi$ and real eigenvalues $\nu$. Using the transform to make the exponentiation faster and then undoing it gives us the final propagator. 
\begin{equation}\label{eq:propagator}
  \mathcal{S}(\tau)\,\alpha=\Pi^{1/2}\Psi
  \Big(e^{\nu\tau}\odot\big(\Psi^{\top}\Pi^{-1/2}\alpha\big)\Big),
\end{equation}
where $\odot$ is the elementwise product and $e^{\nu\tau}$ has entries $e^{\nu_i\tau}$. Only that diagonal factor depends on $\tau$, so applying a propagator costs two matrix--vector products.

Every factor in Eq.~\eqref{eq:oplik} is now written on a discretised grid, and the likelihood follows by chaining them. We drop the dark gaps before the first and after the last photon, since the photons themselves define the start and end of a trace. Applying the killing operator many times over or underflows $\alpha$ quickly. We therefore renormalize $\alpha$ after each photon and accumulate the log-normalizers,
\begin{equation}\label{eq:forward}
  \begin{aligned}
    \tilde{\alpha}_1 &= \Lambda_{c_1}\,\alpha_0 , \qquad
    \tilde{\alpha}_n = \Lambda_{c_n}\,\mathcal{S}(\tau_n)\,\alpha_{n-1}
    \quad (n \geq 2) , \\
    Z_n &= \mathbf{1}^{\top}\tilde{\alpha}_n , \qquad
    \alpha_n = \tilde{\alpha}_n/Z_n ,
  \end{aligned}
\end{equation}
with $\tilde{\alpha}_n$ the filter just after photon $n$ and $Z_n$ its mass, so that $\log L(\boldsymbol{\theta})=\sum_{n=1}^{N}\log Z_n$. The rescaling is exact, because the factors we divide out are the ones we add back in logarithmic form. We implement the recursion in PyTorch~\cite{paszke2019pytorch} and compute batch-independent traces on a single GPU. 

Gradients come from the same computation. Automatic differentiation~\cite{Griewank2008} returns all of them at roughly the cost of one extra evaluation, because every step above is a composition of differentiable operations.

\subsection{Landscape parameterization}
\label{sec:landscape}

We parameterize the free-energy landscape $u(x)$ as a natural cubic spline. The spline has $K$ uniformly spaced knots at positions $x_{\mathrm{s},1},\dots,x_{\mathrm{s},K}$ with spacing $h_{\mathrm{s}}$. The free parameters $\boldsymbol{\mu} = (\mu_1,\dots,\mu_K)^{\top}$ are the heights of the spline function at the knot positions, $u(x_{\mathrm{s},k}) = \mu_k$. The continuous spline function is therefore
\begin{equation}
  u(x) = \sum_{k=1}^{K} \mu_k\, \phi_k(x),
  \label{eq:landscape}
\end{equation}
with $\phi_k$ the natural cubic spline basis functions, which satisfy $\phi_k(x_{\mathrm{s},k}) = 1$ and $\phi_k(x_{\mathrm{s},k'}) = 0$ for $k' \neq k$. This forces the landscape to satisfy $u(x_{\mathrm{s},k}) = \mu_k$, while the coefficients of the cubic splines $\phi_k$ are fixed by the knot geometry alone and do not depend on $\boldsymbol{\mu}$. On a spatial grid $\{x_i\}_{i=1}^{M}$ the evaluation of $u(x)$ can be efficiently expressed as a matrix--vector product $\mathbf{u} = \Phi\boldsymbol{\mu}$, with $\Phi_{ik} = \phi_k(x_i)$. 

\subsection{Optimization}

We aim to infer the parameters $\boldsymbol{\theta}$ that are most compatible with an observed trace $\mathcal{D}$. Therefore, we maximize the objective
\begin{equation}
  \mathcal{J}(\boldsymbol{\theta};\mathcal{D})
  = \log L(\boldsymbol{\theta};\mathcal{D}) + \log p(\boldsymbol{\theta}),
  \label{eq:objective}
\end{equation}
which is the log posterior up to an additive constant. Here  $L(\boldsymbol{\theta}; \mathcal{D})$ is the likelihood written explicitly with the data $\mathcal{D}$ and $p(\boldsymbol{\theta})$ is the prior. Independent traces $\{\mathcal{D}_m\}_{m=1}^{N_{\mathrm{trace}}}$ contribute additively to the log-likelihood  $\log L(\boldsymbol{\theta};\{\mathcal{D}_m\}_{m=1}^{N_{\mathrm{trace}}}) = \sum_{m=1}^{N_{\mathrm{trace}}}\log L(\boldsymbol{\theta};\mathcal{D}_m)$. We minimize $-\mathcal{J}$ with respect to the full parameter vector $\boldsymbol{\theta}$:
\begin{equation}
  \boldsymbol{\theta} = \bigl(\mu_1,\dots,\mu_K,\  D,\  a_g,\  a_r,
  \  \beta_g,\  \beta_r\bigr)^{\top}
\end{equation}
where the $\mu_k$ are the spline knot heights (in $k_{\mathrm{B}}T$) and $K$ is the number of spline knots. $D$ is the diffusion coefficient, and $a_g$, $a_r$, $\beta_g$ and $\beta_r$ are the four identifiable photophysics parameters. $\log p(\boldsymbol{\theta})$ collects all the prior terms, which contain the landscape anchor, curvature prior and background prior (see Sec. \ref{sec:priors}). The parameters that are strictly positive are carried in log space. We use the LBFGS optimizer \cite{liu1989limited} to minimize the negative log posterior. We use a patience based stopping criterion and stop if the likelihood improvement is smaller then 0.1 nats in the last 50 steps (see Appendix \ref{si:fit-params}).

All parameters are initialized directly from the data. The total photon rate of the two channels is set to the observed count rate of the trace, and the background rates to their prior means. The free-energy landscape is initialized from a data-driven estimate from the traces. First we bin the photon stream, compute the FRET efficiency of each bin, and map it onto the reaction coordinate through $E_{\mathrm{obs}}(x) =  \frac{\lambda_r(x)}{\lambda_r(x) + \lambda_g(x)}$ (for $\lambda_c(x)$ see Eq. \eqref{eq:photon_rates}) evaluated at the initial $a_c$ and $\beta_c$. The resulting distribution is smoothed by kernel density estimation with Silverman's bandwidth \cite{silverman2018density} and Boltzmann-inverted to give $u(x)$. The bin size is chosen from a small set of
candidates by held-out log-likelihood. Finally, the diffusion coefficient is initialized by a coarse grid search that maximizes the likelihood with the landscape and photophysical parameters held fixed.

\subsection{Priors}
\label{sec:priors}
We place two prior terms on the landscape. The first enforces local smoothness between knots through a roughness penalty on the discrete second difference,
\begin{equation}
  -\log p_{\mathrm{rough}}(\boldsymbol{\mu})
  = \omega \sum_{k=2}^{K-1}
    \left( \frac{\mu_{k+1} - 2\mu_k + \mu_{k-1}}{h_{\mathrm{s}}^{2}} \right)^{2}.
  \label{eq:rough}
\end{equation}
The second term fixes the landscape offset, which the likelihood leaves undetermined because it depends on $u(x)$ only through energy differences. We anchor the knot mean $\bar\mu = K^{-1}\sum_{k=1}^{K}\mu_k$ with a Gaussian,
\begin{equation}
  -\log p_{\mathrm{anchor}}(\boldsymbol{\mu})
  = \tfrac{1}{2}\bigl(\bar\mu / \sigma_{\mathrm{anch}}\bigr)^{2}
  \label{eq:anchor}
\end{equation}
which selects a single offset without otherwise constraining the shape of the landscape. We set $\sigma_{\mathrm{anch}} = 1\ k_{\mathrm{B}}T$. 
Additionally, we also use a prior for the background rates. We use a Gamma prior for the background $\beta_c$ in each channel. The negative log density of the Gamma distribution can then be expressed up to an additive constant as
\begin{align*}
    -\log p_{\mathrm{background}}(\beta_c) = \frac{\mu_{\beta_c}^2}{\sigma_{\beta_c}^2} \cdot (r -\log r -1)
\end{align*}
with $\beta_c$ the background value in channel $c$ and $r = \beta_c/\mu_{\beta_c}$. The mode of the Gamma distribution is $\mu_{\beta_c}$ and its approximate standard deviation is $\sigma_{\beta_c}$. For all cases, we assume the background is calibrated at the ground truth with a 10\% width.

\subsection{Uncertainty quantification}
\label{sec:uq}
Because the likelihood is differentiable, we quantify parameter uncertainties from the local curvature of the log-likelihood at its maximum. The curvature is the Fisher information $I(\boldsymbol{\theta})$~\cite{LehmannCasella1998} for $N_{\mathrm{trace}}$ independent traces, and its inverse bounds the covariance of an estimator, $\operatorname{Cov}[\hat{\boldsymbol{\theta}}] \succeq I(\boldsymbol{\theta})^{-1}$. Evaluating $I$ from second derivatives would require differentiating the likelihood twice. We can compute it already from first derivatives. For each FRET trace $\mathcal{D}_m$ we compute the score $\mathbf{s}_m = \nabla_{\boldsymbol{\theta}}\log L(\boldsymbol{\theta};\mathcal{D}_m)$ by automatic differentiation, at the cost of one backward pass. At the true parameter, the covariance of the score equals the expected curvature of the negative log-likelihood~\cite{bartlett1953approximate}, and the information is additive over independent traces. We therefore estimate it with the empirical outer-product form~\cite{hall1974estimation},
\begin{equation}
  \hat{I} = \sum_{m=1}^{N_{\mathrm{trace}}} \mathbf{s}_m \mathbf{s}_m^{\top} ,
  \label{eq:fisher}
\end{equation}
where $N_{\mathrm{trace}}$ is the number of independent traces. Evaluated at the MAP rather than the true parameter, this holds as an approximation.

Because we regularise the landscape during optimization, we report the Laplace~\cite{TierneyKadane1986} posterior precision $H$ and its inverse $\Sigma$ as a proxy for the posterior uncertainty,
\begin{equation}
  H = \hat{I}
    + \nabla^{2}_{\boldsymbol{\theta}}\bigl[-\log p(\boldsymbol{\theta})\bigr],
  \qquad
  \Sigma \equiv H^{-1}
    \approx \operatorname{Cov}[\boldsymbol{\theta} \mid \mathcal{D}] .
  \label{eq:laplace}
\end{equation}

We report landscapes with a fixed offset and propagate the covariance, so we report uncertainties which are insensitive to a shift of the landscape. Let $\Sigma_{\boldsymbol{\mu}\boldsymbol{\mu}}$ be the block of $\Sigma$, which contains only the marginal covariances of the knot heights $\boldsymbol{\mu}$. We fix the shift of the spline function $u(x)$ simply by fixing the shift  of the basis functions on the analysis grid $\{x_i\}_{i=1}^{M}$, with $ \bar\phi_k(x) = \phi_k(x)
    - \frac{1}{M}\sum_{i=1}^{M} \phi_k(x_i) $
    
The pointwise uncertainty of the landscape can then be computed by propagating the knot covariance through the centred basis 
\begin{equation}
  \sigma_u^{2}(x)
   = \sum_{k,k'=1}^{K} \bar\phi_k(x)\,
     \bigl[\Sigma_{\boldsymbol{\mu}\boldsymbol{\mu}}\bigr]_{kk'}\,
     \bar\phi_{k'}(x) ,
  \label{eq:sigma_u}
\end{equation}
which on the grid collapses to $\operatorname{diag}\bigl(\bar\Phi\,
\Sigma_{\boldsymbol{\mu}\boldsymbol{\mu}}\,\bar\Phi^{\top}\bigr)$ with $\bar\Phi_{ik} = \bar\phi_k(x_i)$. Since $\sum_k \phi_k(x) = 1$, the centered
basis satisfies $\sum_k \bar\phi_k(x) = 0$, so Eq.~\eqref{eq:sigma_u} is insensitive to a uniform shift of the knot heights. For the error of the barrier height $\sigma_{\Delta u^\ddagger}$ we can use the same approach. We first define the vector $g_k = \phi_k(x^{\ddagger}) - \phi_k(x_{min})$ for $k = 1, .., K$, with $x_{min}$ the position of the well minimum and $x^{\ddagger}$ the position of the barrier top. We can compute the barrier height from the knot positions with $\Delta u^\ddagger = \boldsymbol{g}^\top\boldsymbol{\mu}$, which gives us the uncertainty with  $\sigma_{\Delta u^\ddagger}^2 = \boldsymbol{g}^\top\Sigma_{\mu\mu}\boldsymbol{g}$.

The diffusion coefficient and the four photophysics parameters carry no landscape offset. For these we report the marginal posterior standard deviation $\sigma_{\log D} = \sqrt{[\Sigma]_{\log D,\log D}}$, mapped to physical units by the delta method, $\sigma_D = \hat{D}\,\sigma_{\log D}$ and likewise for each rate.

\subsection{Forward model and synthetic data}
\label{sec:forward}
We generate synthetic photon streams by simulating the donor-acceptor distance $x(t)$ with overdamped Langevin dynamics and then simulating the FRET photophysics. 
We simulate $x(t)$ by using the Euler–Maruyama integration scheme~\cite{KloedenPlaten1992}
\begin{equation}
  \begin{aligned}
    x_{j+1} &= x_j-D\,u'(x_j)\,\Delta t +\sqrt{2D\Delta t}\;\xi_j.
  \end{aligned}
  \label{eq:euler}
\end{equation}
with $\xi_j\sim\mathcal{N}(0,1)$. $u(x)$ is a natural cubic spline, so the force is evaluated analytically from the spline derivative $u'(x)$. $\Delta t$ is the integration time step and each trajectory starts from the equilibrium density $x_0\sim\pi_{\boldsymbol{\theta}}\propto e^{-u^(x)}$.
 Since $\lambda_c$ depends only on $x(t)$, the photon stream is an inhomogeneous Poisson process: within step $j$ we draw $\Delta N_c\sim\mathrm{Poisson}(\lambda_c(x_j)\Delta t)$ and place arrivals uniformly on $[j\Delta t,(j{+}1)\Delta t)$, yielding a marked point process $\{(t_n,c_n)\}$ of arrival times and channel labels.

\section{Results}
\label{sec:results}
In this section we showcased our method on simulated data. We recovered the free-energy landscape and the diffusion coefficient for systems ranging from a two well system to more complex landscapes with a short-lived intermediate. We also benchmarked the speed of the implemented likelihood and showed how it can be use to guide the design of new experiments.

\begin{figure}
\centering
\includegraphics[width=\columnwidth]{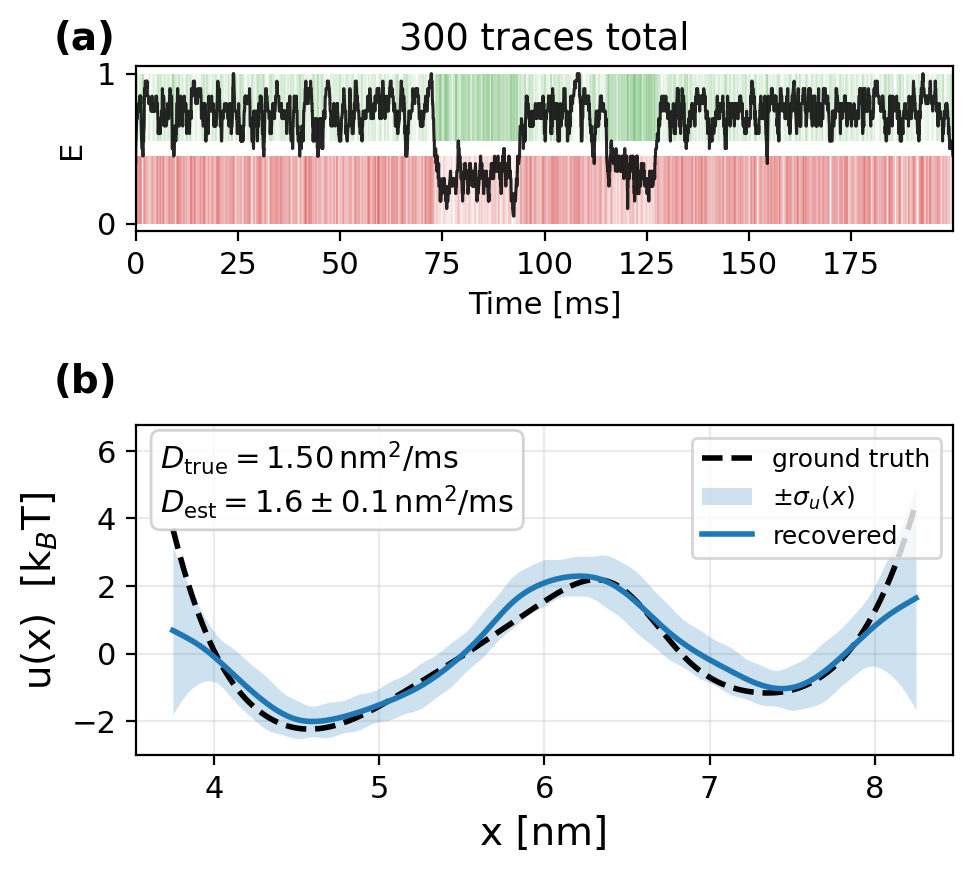}
\caption{\label{fig:fig2}%
\textbf{Inference of an asymmetric two-state landscape.} Inference on simulated
data for a two-state system with an asymmetric barrier and realistic
photophysics. (a) Representative smFRET photon stream, with estimated FRET efficiency E (black, one of $300$ traces). FRET efficiencies were obtained by binning the photon stream in 2 $\mathrm{ms}$ windows and computing the fraction of acceptor photons, in each bin. The vertical ticks mark individual photons, colored
by detection channel. (b) Recovered landscape $\hat u(x)$ (blue) with $\pm\sigma_{u}(x)$
uncertainty band, against the ground truth (black, dashed). The diffusion coefficient is
inferred jointly with $\hat u(x)$.}
\end{figure}

To test our framework we started by producing synthetic FRET traces for which we know the ground truth parameters $\boldsymbol{\theta}^{*}$. The ground truth parameters describe a free-energy landscape $u^*(x)$ that has two wells separated by an asymmetric barrier of around 5 $k_{\mathrm{B}}T$ with a diffusion coefficient of $D^{*} = 1.5\ \mathrm{nm^2\,ms^{-1}}$. Additionally we used realistic brightness, background, and crosstalk (see Appendix \ref{si:sim-params}) and set $R_0= 6\ \mathrm{nm}$. We simulated  $N_{\mathrm{trace}} = 300$ independent smFRET traces (Fig.~\ref{fig:fig2}a)  from the ground truth parameters, with the total dataset $\{\mathcal{D}_m\}_{m=1}^{N_{\mathrm{trace}}}$containing approximately 1.5 million photons in total.

We aimed to infer the parameters $\boldsymbol{\theta}$ that best describe the data $\{\mathcal{D}_m\}_{m=1}^{N_{\mathrm{trace}}}$. Therefore, we performed gradient-based optimization of the unnormalized log-posterior $\mathcal{J}(\boldsymbol{\theta}\,;\,\{\mathcal{D}_m\}_{m=1}^{N_{\mathrm{trace}}})$. Compared to the raw likelihood $L$, the log-posterior $\mathcal{J}$ additionally includes a prior. The prior comprises three terms: a smoothness penalty on the free-energy landscape, a constraint on the offset, and priors on the background rates. Our best estimate of the parameters sits at the minimum of the negative log-posterior
\begin{equation}\label{eq:mle}
  \hat{\boldsymbol{\theta}}
  =\operatorname*{arg\,min}_{\boldsymbol\theta} -\mathcal{J}(\boldsymbol{\theta};\{\mathcal{D}_m\}_{m=1}^{N_{\mathrm{trace}}})
\end{equation}
The optimal strength $\omega$ of the smoothness penalty is not known a apriori,  we therefore scanned a range of values and kept the fit that maximizes the log-likelihood on a set of traces not used for optimization (see Fig. S1a). The inferred parameters $\hat{\boldsymbol{\theta}}$ recovered both the ground-truth landscape $u^{*}(x)$ and the diffusion coefficient $D^{*}$ (Fig.~\ref{fig:fig2}b). Additionally, the four identifiable photophysics parameters $a_g, a_r, \beta_g, \beta_r$ are all estimated to within 3\% of the ground truth.
 
The inferred parameters $\hat{\boldsymbol{\theta}}$ provide us with a point estimate, but do not contain any information about the uncertainty.  The differentiability of the likelihood not only helps us with the optimization, but it also provides us with a straightforward way to estimate  the uncertainty. By looking at the local curvature of the posterior around the estimated parameters $\hat{\boldsymbol{\theta}}$, we can estimate the uncertainty. A steep, narrow optimum pins the estimate down tightly and implies a small uncertainty, whereas a flat and shallow optimum leaves the parameter poorly constrained and implies a large uncertainty. The local curvature of the likelihood is defined by the Fisher information matrix $\hat{I}(\hat{\boldsymbol{\theta}})$. For the posterior curvature, including the information from the prior, we denote it as $H$. The inverse $H^{-1}$ is a proxy for the parameter covariance around $\hat{\boldsymbol{\theta}}$, and provides us with $1\sigma$ confidence interval of each parameter (see Sec.\ref{sec:methods}). Figure~\ref{fig:fig2}b shows the $1\sigma$ intervals for the free-energy landscape $\hat{u}(x)$ and for $\hat{D}$. The uncertainty band $\sigma_{u}(x)$describes how much the shape of the free-energy landscape can change within the $1\sigma$ uncertanty. The confidence was not uniform along the donor-acceptor distance $x$: the interval widened wherever $E(x)$ is flat. In the regions where a change of $x$ barely shifts the FRET efficiency, the data constrained the landscape only weakly.

\begin{figure}
\centering
\includegraphics[width=\columnwidth]{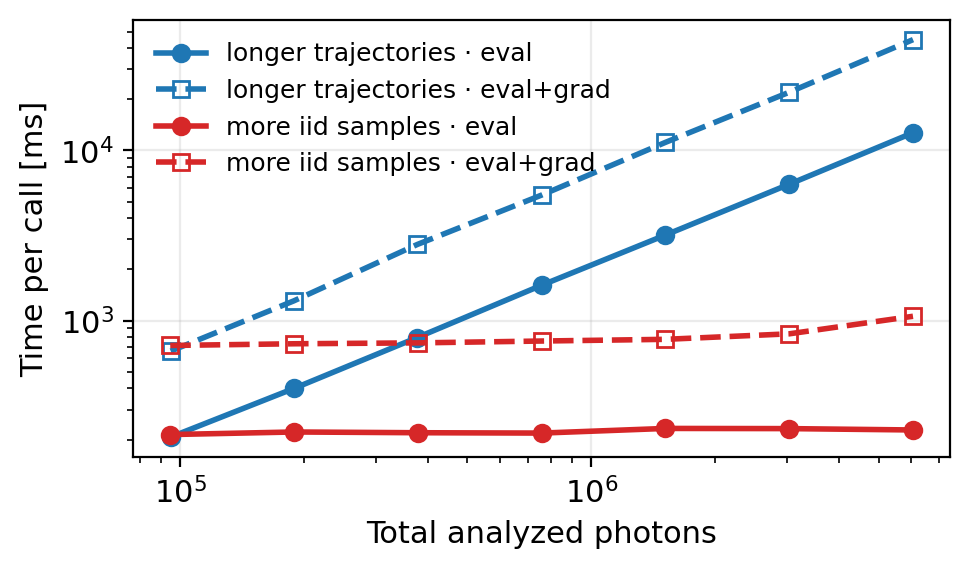}
\caption{\label{fig:fig3}%
\textbf{Computational scaling of the likelihood.} Wall-clock time per likelihood
call versus the total number of analyzed photons, for evaluation alone
(\emph{eval}, solid) and evaluation with its gradient (\emph{eval+grad}, dashed).
Adding photons by lengthening a single trajectory (blue) scales approximately
linearly, while adding photons by analyzing more independent traces (red) scales much slower. The first evaluation point at $10^5$ photons is setup for both cases, at every other point the traces are added or extended.}
\end{figure}

The likelihood reduces to dense, automatically differentiable linear algebra, so both the likelihood and its exact gradient evaluate efficiently on a GPU. To systematically assess the evaluation speed, we benchmarked the wall-clock cost of a single likelihood call, with and without gradient evaluation, as a function of the total number of photons analyzed (Fig.~\ref{fig:fig3}).  We used the setup from the two-well system above and either change the total simulation time per trace or the number of independent traces, to change  the number of analyzed photons. 
How that computational cost grows depends on how the photons are arranged. All the computations for a single trace need to be performed in sequence, each propagator in Eq. (\ref{eq:oplik}) waiting for the previous one, so the cost grows approximately linearly. Therefore, analysing datasets where data are added by lengthening the traces resulted in a linear increase in the computation time per likelihood call (Fig. \ref{fig:fig3}, blue lines).  On the other hand, the likelihood computation for many independent traces can be efficiently parallelized. That means on a GPU we can compute the likelihood for many traces together at once. So the time for analysing datasets where data is added by adding independent traces grew much more slowly (Fig. \ref{fig:fig3}, red line), and was mainly limited by GPU memory.

Evaluating the gradient multiplied the cost by a roughly constant factor. To set these numbers in context, a full optimization run at $\sim\!10^{6}$ photons completed in 3 min on a single RTX A6000 GPU. For a fixed photon budget, photons spread over independent traces are far cheaper to analyse than photons concentrated in one long trajectory.

We can also use the likelihood to quantify how much a proposed experiment can reveal before that experiment is run. Because parameter uncertainty follows quickly from the curvature of the likelihood, we can assess how tightly the ground-truth parameters are constrained under an assumed setup. 

\begin{figure}[b]
\centering
\includegraphics[width=\columnwidth]{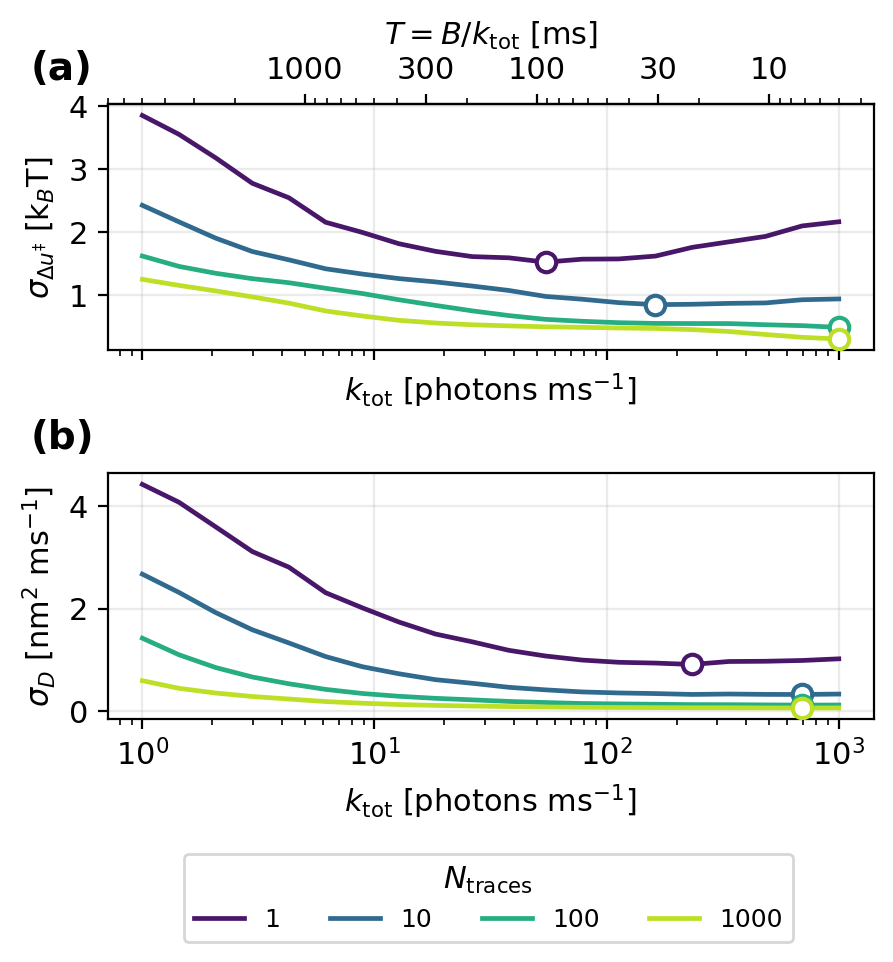}
\caption{\textbf{Uncertainty depends on how photons are acquired.} Predicted $1\sigma$ uncertainties on (a)~the diffusion coefficient $D$ and (b)~the barrier height $\Delta u^{\ddagger} \equiv u(x^{\ddagger}) - u(x_{\mathrm{min}})$ at the ground-truth parameters $\theta^{*}$. With $x^\ddagger$ the position of the barrier top and $x_{\mathrm{min}}$ the position of the well minima. Each molecule yields a fixed budget of $B = 5000$ detected photons, so the detection rate $k_{\mathrm{tot}}$ (bottom axis) sets the trace duration $T = B/k_{\mathrm{tot}}$ (top axis). Colors denote the number of molecules $N_{\mathrm{trace}}$ and the open circles mark each curve's optimal $k_{\mathrm{tot}}$. Uncertainties come from the posterior curvature $H$ evaluated at $\theta^{*}$ [Eq.~\eqref{eq:laplace}], so the figure can be computed before any measurement.}
\label{fig:fig4}
\end{figure}

Photobleaching is the light-induced destruction of a FRET dye, it is the main constraint on the number of photons available in a single-molecule FRET experiment. As a simple bleaching model, we assume that a dye emits a finite number of photons $B$ before it bleaches. The laser intensity therefore does not set how many photons a molecule yields, but only how quickly it yields them. We quantify the laser intensity by the total detection rate $k_{\mathrm{tot}}$, the mean rate at which we detect photons in both channels, including the background and the detection efficiencies. Choosing $k_{\mathrm{tot}}$ fixes the trace duration at $T = B / k_{\mathrm{tot}}$. The two extreme acquisition strategies are therefore bright and short traces, or dim and long traces. Lets for now consider the setup from the two-well system above.  To find where between these extremes an experiment is most informative, we fixed $B = 5000$ photons per molecule and scanned $k_{\mathrm{tot}}$ logarithmically over three orders of magnitude, from $1$ to $10^{3}$ photons per millisecond. The trace duration then ran from $5\ \mathrm{s}$ down to $5\ \mathrm{ms}$. We set $k_{\mathrm{tot}}$ at each point by rescaling the excitation and background rates by a common factor. Every point therefore collected the same number of photons per molecule, at the same background fraction, and differed only in how those photons were spread over time. We simulated a set of photon traces at each point and evaluated the uncertainty our likelihood assigns to the parameters at the ground truth (see Sec.~\ref{sec:uq}). The independent choice is how many molecules $N_{\mathrm{trace}}$ to record sets the total photon budget $N_{\mathrm{trace}} B$. Figure~\ref{fig:fig4} reports the resulting uncertainty of the diffusion coefficient $D$ and of the barrier height $\Delta u^{\ddagger}$ for several values of $N_{\mathrm{trace}}$.
 
For a single trace, the uncertainty on the barrier height showed a clear optimum. At low $k_{\mathrm{tot}}$ the trace is long and the molecule crosses the barrier many times, but the photon rate is too low to resolve any single one of them. At high $k_{\mathrm{tot}}$ every trajectory is sharply resolved, but the trace is too short to contain enough crossings. The barrier height is best constrained where the two effects balance. The uncertainty on the diffusion coefficient behaved differently. It decreased with $k_{\mathrm{tot}}$ and then saturated, without an interior optimum. $D$ is informed by the short-time dynamics, so brighter traces are always preferable, and nothing is gained once the photon spacing resolves the fastest motion. Increasing $N_{\mathrm{trace}}$ lowered both uncertainties and shifted the optimum of $\sigma_{\Delta u^{\ddagger}}$ towards shorter and brighter traces. For many traces the optimum moved to the bright end of the sweep. This is possible because each trace is initialized from the Boltzmann distribution. The ensemble then explores the landscape in parallel, and no single trace has to be long enough to cross the barrier repeatedly. The location of the optima also depends on the dynamics of the system, and faster dynamics require a higher $k_{\mathrm{tot}}$ (see Fig. S2). The two parameters therefore do not share a common optimum, and the uncertainty on $D$ favors a higher flux than the uncertainty on $\Delta u^{\ddagger}$. The best acquisition strategy depends on the target and on the timescales of the system. But with the likelihood this choice can be quantified before any data are recorded.

In a last, more difficult test, we added a weakly populated third state, a short-lived intermediate the molecule visits only in passing between the two main states.  The molecule visits the intermediate only briefly and together with the realistic background it did not form its own peak in the FRET-efficiency histogram (see Fig. S3)
We started by simulating 300 traces $\{\mathcal{D}_m\}_{m=1}^{N_{\mathrm{trace}}}$ with a free-energy landscape which contains a short-lived intermediate and a diffusion coefficient of $D^{*} = 1.5\ \mathrm{nm^2\ ms^{-1}}$ (Fig. \ref{fig:fig5}a). The photophysics were set to realistic brightness, background, and crosstalk (see Appendix \ref{si:sim-params}), and we set $R_0 = 5.5\ \mathrm{nm}$. In total the dataset $\{\mathcal{D}_m\}_{m=1}^{N_{\mathrm{trace}}}$ contained around 1.5 million photons.  We inferred the underlying parameters $\hat{\boldsymbol{\theta}}$ from $\{\mathcal{D}_m\}_{m=1}^{N_{\mathrm{trace}}}$ using gradient based optimization (see Fig. S1b, same selection strategy for the smoothness penalty $\omega$ as before). From the estimated parameters $\hat{\boldsymbol{\theta}}$, we recovered the shape of the underlying free-energy landscape (Fig. \ref{fig:fig5}). The short-lived intermediate appeared clearly as a third state in the free-energy landscape.

We repeated this experiment 50 times with independently simulated data sets  $\{\mathcal{D}_m\}_{m=1}^{N_{\mathrm{trace}}}$ (Fig. \ref{fig:fig5} gray lines). We can see that, we recovered the intermediate in every case. Additionally, this shows that the estimated uncertainties are well calibrated, as the uncertainty of a single estimate explained the spread over multiple independent datasets. The same calibration was observed for $D$ (see Fig. S4).

\begin{figure}
\centering
\includegraphics[width=\columnwidth]{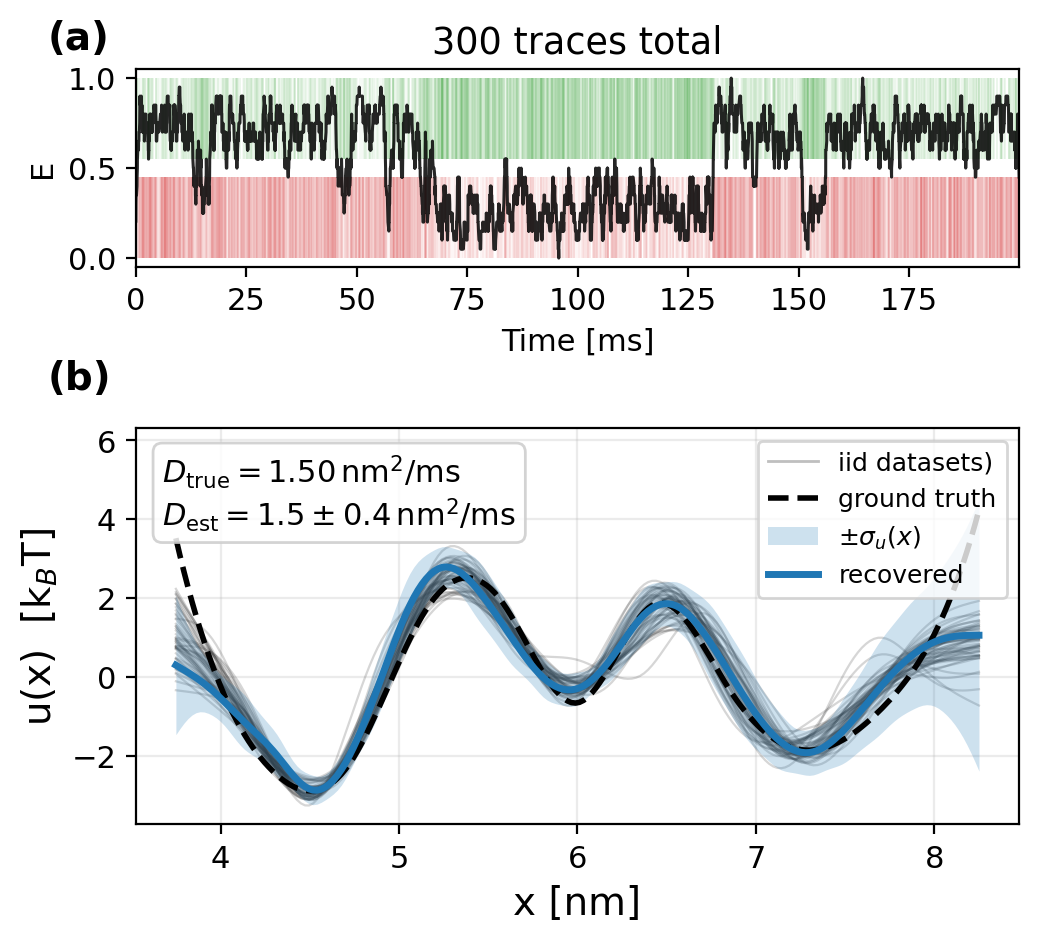}
\caption{\label{fig:fig5}%
\textbf{Inference for a intermediate state : } Inference on simulated data
for a landscape with a weakly populated intermediate and realistic photophysics.
(a) Representative trace with estimated FRET efficiency $E$ (black, ; one of
$300$ traces). The FRET efficiencies E were obtained by binning the photon stream in 2 $\mathrm{ms}$ windows and computing the fraction of acceptor photons, in each bin. The vertical ticks mark individual photons, colored by detection channel. (b) Recovered landscape $\hat u(x)$ (blue) with $\pm\sigma_u(x)$ uncertainty band, against the ground truth (black, dashed). The diffusion coefficient is inferred jointly with $\hat u(x)$. Repeated inference with independently simulated data sets recovers the same free energy landscape within the predicted uncertainty (gray lines)}.
\end{figure}

Recovering the ground-truth parameters is reassuring but cannot be performed in any setting with real experimental data.  In this case, we can ask whether the inferred model reproduces the measured photon streams themselves. We therefore performed predictive checks. Therefore,  we generated a fresh set of 300 simulations  $\{\mathcal{D}^{\mathrm{sim}}_m\}_{m=1}^{N_{\mathrm{trace}}}$ from the estimated parameters $\hat{\boldsymbol{\theta}}$ and then compared them against simulations from the ground truth. 

\begin{figure}
\centering
\includegraphics[width=\columnwidth]{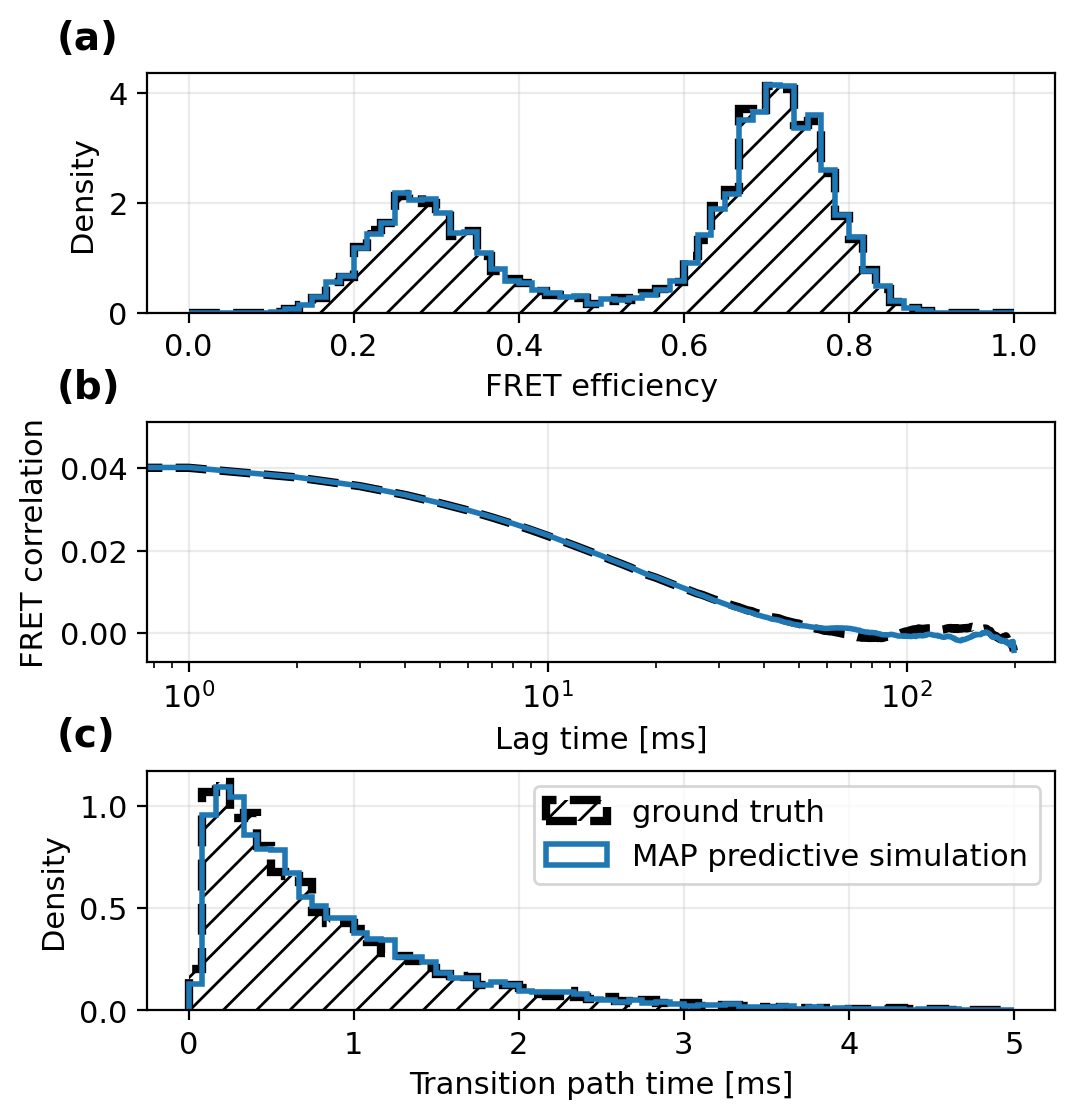}
\caption{\label{fig:fig6}\textbf{Predictive checks.} Simulations generated using the estimated parameters (blue) are compared against the ground-truth simulations (black). (a) FRET efficiency histograms (2 $\mathrm{ms}$ bin width), probing the equilibrium populations. (b) FRET correlation functions, probing the trajectory dynamics. (c) Transition path time distributions, computed from the ground-truth and estimated parameters.}
\end{figure}

First, we compared the FRET-efficiency histogram, which reflects the equilibrium populations. We can see that the simulation $\{\mathcal{D}^{\mathrm{sim}}_m\}_{m=1}^{N_{\mathrm{trace}}}$ from the inferred parameters $\hat{\boldsymbol{\theta}}$  and the simulation from the ground truth parameters $\boldsymbol{\theta}^{*}$ produced matching efficiencies histograms (Fig.~\ref{fig:fig6}a). As a second measure, we compared the FRET correlation function \cite{terterov2025model}, which is informative about the dynamics (Fig.~\ref{fig:fig6}b). The FRET correlation function also matched very well. These checks one could run on experimental data, where the true parameters are unknown. 

Finally, the fitted model also gives us access to quantities that are not contained in the parameters themselves. One important example is the distribution of transition path times, which is notoriously hard to measure experimentally \cite{sturzenegger2018transition, neupane2016direct, chung2012single}. But from a free-energy profile and a diffusion coefficient it can be computed directly. Therefore, we simulated  many barrier crossings and record their durations (see Appendix \ref{si:TPS}). We do this for both the ground-truth and the estimated parameters, and the two transition path time distributions agreed closely (Fig.~\ref{fig:fig6} c). Our inference therefore recovers not only the parameters of the model but also the dynamical observables that follow from them. Applied to measured photon streams, this offers an indirect route to transition path times.

The photon stream of an individual trace also constrains what a particular molecule did, and the likelihood already contains this information. Equation~\eqref{eq:oplik} weights every trajectory by its probability under Eq.~\eqref{eq:langevin} and by the conditional likelihood of Eq.~\eqref{eq:cond-lik}. Integrating the position out at the end collapses this to the single number $L(\boldsymbol{\theta})$. Keeping it instead gives the posterior $p\big(x(t)\mid\mathcal{D},\hat{\boldsymbol{\theta}}\big)$ over the hidden distance. A second sweep through the trace, propagated backwards from $T$, gives us this posterior at every time $t$. From it we can get the exact posterior mean, and exact sampled paths, at full time resolution (see Appendix \ref{si:reconstruction}).

\begin{figure}
\centering
\includegraphics[width=\columnwidth]{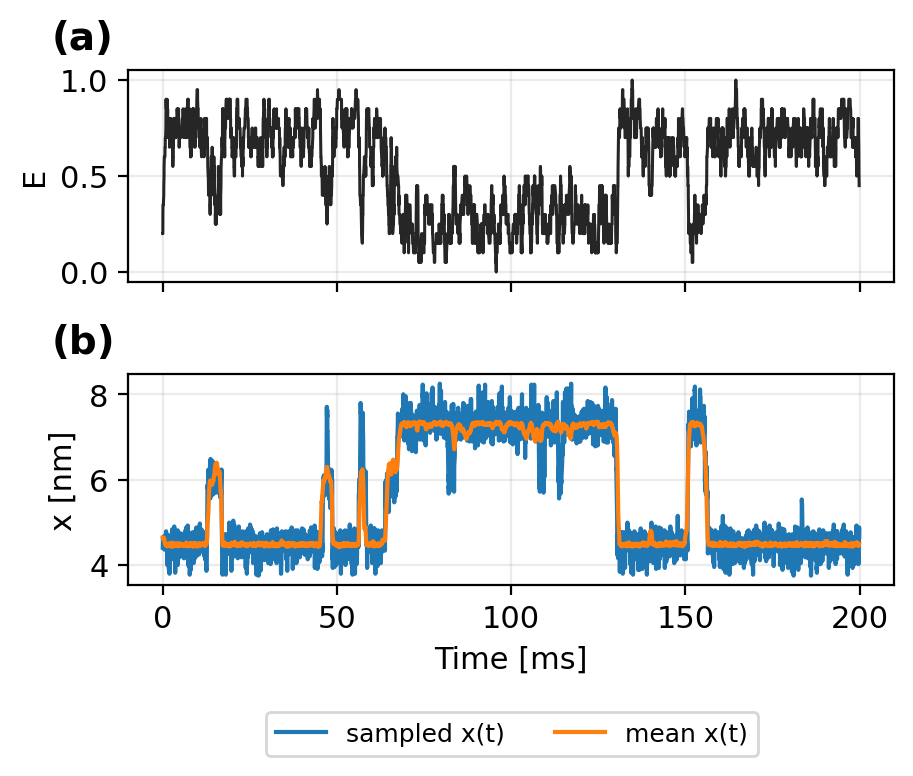}
\caption{\textbf{Reconstructing the distance trajectory from a photon stream.}
(a) FRET efficiency E from a single trace, obtained by binning the simulated photon stream in windows of  2 $\mathrm{ms}$ and computing the fraction of acceptor photons per bin. (b) Posterior
mean of the donor-acceptor distance given the complete photon stream (orange), together with one sampled trajectory from the posterior (blue).}
\label{fig:fig7}
\end{figure}

Figure~\ref{fig:fig7} shows the result for a simulated photon stream (Fig. \ref{fig:fig7}a), here from a landscape with an additional intermediate state. We plotted the posterior mean (Fig. \ref{fig:fig7}b, orange) together with one sample (Fig. \ref{fig:fig7}b, blue) from the posterior over paths. The mean tracked the FRET efficiencies and we obtained the mean position at every time. Therefore the mean path resembled what a state-based analysis would return: flat stretches joined by fast transitions. That resemblance is a consequence of averaging and not an assumption. Every trajectory consistent with the photons enters the average, the fast diffusive fluctuations differ between them and get cancel out, and what remains is the part they agree on. The sample shows what the cancellation removes. A diffusive trajectory is rough, so the mean is not a typical path. 

The two analyses answer different questions. Simulating transitions with the fitted model gives the distribution of transition path times across the ensemble. The reconstruction returns the trajectories a particular photon stream is consistent with.

\section{Discussion}
Characterizing conformational dynamics in biomolecules remains difficult. Single-molecule FRET probes these dynamics, but only indirectly, through a stream of colored photons. Recovering the free-energy landscape and the diffusion coefficient from such a stream of photons is a hard inverse problem. We addressed it by deriving an exact, differentiable photon-by-photon likelihood for one-dimensional overdamped diffusion on a continuous free-energy landscape.

On simulated FRET data, our inference recovers the free-energy landscape, the diffusion coefficient, and the photophysical parameters together, without fixing the number of conformational states in advance. We quantify the uncertainty of these estimates directly from the curvature of the likelihood. The approach also held up on a harder problem. A short-lived
intermediate that the molecule visits only passing between the two main states, yet our inference still recovered it as a distinct third well in the landscape.

The likelihood is also cheap to evaluate. A single evaluation and its gradient stay inexpensive as the number of photons grows. This matters because computational cost currently limits how far inference with flexible models can be scaled to the large and heterogeneous datasets now common in the field \cite{smfret_review}. Our analyses used about $1.5\times10^{6}$ photons in total, a size routinely reached in a FRET experiment. The curvature of the likelihood also tells us, before any data are collected, how well a given setup can resolve the parameters. This is a first step towards designing experiments that resolve the underlying parameters of interest optimally.

Our model rests on several assumptions. We reduce the dynamics to a single reaction coordinate, and we assume overdamped Langevin dynamics with a constant diffusion coefficient and no memory effects. Some of these assumptions can be tested rather than trusted. Triplet blinking can be checked for beforehand, and non-Markovianity leaves a signature in the FRET correlation function that a predictive check can detect \cite{terterov2025model, satija2019generalized,  berezhkovskii2020nonequilibrium}. Others can be relaxed within the same likelihood. Position-dependent diffusion enters directly into the Fokker-Planck equation and could itself be inferred \cite{best2010coordinate, hummer2005position}. Memory effects could be incorporated by solving the Fokker-Planck equation on a coupled two-dimensional surface whose projection onto a one-dimensional coordinate is non-Markovian. Triplet blinking can likewise be added directly. Each of these extensions costs more parameters, and therefore more data.

Transitioning from the point estimates to a  fully Bayesian answer requires no new machinery. The gradients that drive the optimization are exactly the ones needed by Hamiltonian Monte Carlo sampler. So sampling the full posterior is a direct extension at the cost of longer compute. This would replace the the error estimates with exact posteriors. For dynamics whose likelihood cannot be written down, the same posterior could instead be targeted from synthetic data alone using simulation-based inference, as it has been shown for free-energy landscapes from  force spectroscopy \cite{dingeldein2023simulation, dingeldein2026pred, dingeldein2025simulation}.

Inferring dynamical models from smFRET data is important for understanding complex biomolecular systems\cite{galvanetto2023extreme}. By fitting a dynamical model directly photon-by-photon, our likelihood uses the correlations in photon color and timing that binning discards, and extracts more information from the same data. Being exact and differentiable, it reconstructs full free-energy landscapes and diffusion coefficients with uncertainties using only gradient-based optimization. This puts inference of flexible dynamical models within reach for the smFRET community.

\begin{acknowledgments}
L.D. and R.C. acknowledge the support of Goethe University Frankfurt, the Frankfurt Institute for Advanced Studies, the LOEWE Center for Multiscale Modelling in Life Sciences of the state of Hesse, the CRC 1507: Membrane-associated Protein Assemblies, Machineries, and Supercomplexes (P09), and the Deutsche Forschungsgemeinschaft (DFG, German Research Foundation) under Germany's Excellence Strategy -- EXC 3094 -- 533751785, as well as computational resources and support from the Center for Scientific Computing of Goethe University and the J\"ulich Supercomputing Centre. We were assisted by Claude (Opus 4.8, Opus 5, and Fable 5; Anthropic).
\end{acknowledgments}

\section*{Data Availability Statement}
The code is based on Pytorch\cite{paszke2019pytorch} and is available at GitHub \url{https://github.com/covinolab/diff_fret_likelihood}. Data and scripts necessary to reproduce all the results presented in this paper are freely accessible at the Zenodo repository \url{https://zenodo.org/record/22044845}. We also provide a self contained tutorial where you can use the code yourself in a google Colab notebook \url{https://github.com/covinolab/diff_fret_likelihood/blob/main/tutorial/smfret_likelihood_tutorial.ipynb}.

\bibliography{references.bib}

\appendix
\section{Simulation parameters}
\label{si:sim-params}
\begin{table}[H]
\caption{\label{tab:sim-params}%
\textbf{Parameters of the synthetic smFRET simulator.}  Where the two test systems differ we quote two-state\,/\,three-state.}
\begin{ruledtabular}
\begin{tabular}{llll}
Symbol & Value & Unit & Meaning \\
\colrule
\multicolumn{4}{l}{\textit{Langevin dynamics}} \\
$D^{*}$          & $1.5$              & nm$^2$\,ms$^{-1}$ & Diffusion coefficient \\
$\Delta t$       & $5 \times 10^{-6}$ & ms                & Integration time step \\[2pt]
\multicolumn{4}{l}{\textit{Photophysics}} \\
$R_0$            & $6.0$ / $5.5$      & nm                & F\"orster radius \\
$k_{\mathrm{em}}$           & $30.0$             & ms$^{-1}$         & Donor emission rate \\
$\eta_g$         & $0.8$              &                   & Green detection efficiency \\
$\eta_r$         & $0.8$              &                   & Red detection efficiency \\
$\beta_g$        & $1.6$              & ms$^{-1}$         & Green background rate \\
$\beta_r$        & $4.0$              & ms$^{-1}$         & Red background rate \\
$C_{g\to g}$     & $0.97$             &                   & Donor photon into green \\
$C_{r\to r}$     & $0.92$             &                   & Acceptor photon into red \\
$C_{g\to r}$     & $0.03$             &                   & Donor leakage into red \\
$C_{r\to g}$     & $0.08$             &                   & Acceptor leakage into green \\[2pt]
\multicolumn{4}{l}{\textit{Acquisition}} \\
$T$              & $200.0$            & ms                & Trace duration \\
$N_{\mathrm{trace}}$           & $300$              &                   & Number of independent traces \\
\end{tabular}
\end{ruledtabular}
\end{table}

\section{Transition path simulations}
\label{si:TPS}
We sample the transition path ensemble directly, using a Doob $h$-transform. For overdamped Langevin dynamics on the spline potential $u(x)$, the committor between two reference states $a$ and $b$ satisfies $q'(x) \propto e^{u(x)}$, so that
\begin{equation}
  q(x) = \frac{\int_a^x e^{u(y)} \, \mathrm{d}y}{\int_a^b e^{u(y)} \, \mathrm{d}y} .
  \label{eq:committor}
\end{equation}
Conditioning the dynamics on reaching $b$ before returning to $a$ adds a drift $2 D \, q'(x)/q(x)$, which gives the reactive process

\begin{equation}
  x_{j+1} = x_j
          + \left[ -D\, u'(x_j) + 2 D\, \frac{q'(x_j)}{q(x_j)} \right] \Delta t
          + \sqrt{2 D\, \Delta t}\; \xi_j ,
  \label{eq:doob-em}
\end{equation}
with $\xi_j$ sampled from a Gaussian with zero mean and unit variance.  Every trajectory of this process is reactive, so each one yields a transition path time, and no computation is spent on unproductive recrossings.

We evaluate the committor once on a uniform grid by trapezoidal integration of $e^{u}$.  We locate the two isocommittor interfaces by linear interpolation of the cumulative integral, start each path on the $q = \epsilon$ interface, and integrate with an Euler--Maruyama scheme until the path reaches $q = 1 - \epsilon$.  The lower interface is reflecting, which keeps the walker away from the $q \to 0$ singularity of the drift, and the upper interface is absorbing.  We set $\epsilon = 0.05$.

\section{Hyperparameters}
\label{si:fit-params}
\begin{table}[H]
\squeezetable
\caption{\label{tab:fit-params}%
\textbf{Hyperparameters of the inference pipeline.}  Symbols in
the right-hand column refer to Sec.~\ref{sec:methods} of the main text.}
\begin{ruledtabular}
\begin{tabular}{llp{0.38\columnwidth}}
Parameter & Value & Meaning \\
\colrule
\multicolumn{3}{l}{\textit{Grid and spline model}} \\
\texttt{n\_grid}                 & $200$          & Spatial grid points $M$ \\
\texttt{n\_knots}                & $25$    & Spline knots $K$ of $u(x)$ \\
\texttt{min\_x}, \texttt{max\_x} & $[3.75, 8.75]$ & Spatial grid limits (nm) \\
\multicolumn{3}{l}{\textit{Optimiser (LBFGS)}} \\
\texttt{steps}             & $500$ & Max Iterations\\
\texttt{lr}                & $1$         & Learning rate \\
\texttt{history\_size} & 100 & LBFGS history size \\
\texttt{stop\_patience} & 50 & Stopping patience \\
\texttt{stop\_min\_delta} & 0.1 & Stopping improvement\\
\multicolumn{3}{l}{\textit{Prior and regularisers}} \\
\texttt{curvature\_weight}       & $2.15\cdot 10^{-4}$      & Roughness weight $\omega$ \\
\texttt{bg\_g\_mode} & 1.6 & $\mu_{\beta_g}$\\
\texttt{bg\_r\_mode} & 4.0 & $\mu_{\beta_r}$ \\
\texttt{bg\_g\_sd} & 0.16 & $\sigma_{\beta_g}$ \\
\texttt{bg\_r\_sd} & 0.4 & $\sigma_{\beta_r}$ \\
\texttt{gauge\_sd}               & $1.0$          & Anchor width $\sigma_{\mathrm{anch}}$ ($k_{\mathrm B}T$) \\
\end{tabular}
\end{ruledtabular}
\end{table}

\section{Posterior over the hidden distance trajectory}
\label{si:reconstruction}
We report the posterior mean and posterior samples of the donor--acceptor
distance given a measured photon stream.  Throughout this section the
parameters are held fixed at the estimate $\hat{\bm\theta}$ and dropped from
the notation.
 
\subsection{Backward filter}
The forward filter of Eq.~\eqref{eq:filter} is the joint density of the position $x$ and the photons recorded up to time $t$.  The backward filter is the probability of the photons recorded after $t$, as a function of the position $x$ and time $t$
\begin{equation}
  \chi(x,t) = p\big(\mathcal{D}_{(t,T]}\,\big|\,x(t)=x\big) ,
  \label{eq:backfilter}
\end{equation}
Here $\mathcal{D}_{(t,T]}$ denotes the photons from time $t$, to the end $T$. The two filters are different kinds of objects: $\rho$ is a density over the position, carried jointly with the past record, whereas $\chi$ is the probability of the future record as a function of the position.  At $t=T$ nothing remains to be explained, so $\chi(x,T)=1$.  Repeating the infinitesimal argument that led to Eqs.~\eqref{eq:killedFP} and \eqref{eq:jump}, but running it backwards from
$T$, gives
\begin{equation}
  -\partial_t\chi = \big(\mathcal{L}^{\dagger}-\lambda_{\mathrm{tot}}\big)\chi ,
  \qquad
  \chi(x,t_n^{-}) = \lambda_{c_n}(x)\,\chi(x,t_n^{+}) ,
  \label{eq:backward_pde}
\end{equation}
with $\mathcal{L}^{\dagger}f=D(\partial_x^{2}f-u'\partial_x f)$ the adjoint of Eq.~\eqref{eq:generator}. Both filters propagate through the same dynamics but look in opposite directions in time: $\rho$ at one point collects the probability arriving from every possible previous position, while $\chi$ at a point averages over every possible next position.  Swapping previous and next turns $\mathcal{L}$ into $\mathcal{L}^{\dagger}$, which on the grid, literally means the transpose in Eq.~\eqref{eq:backward_grid}.
 
\subsection{Smoothing marginals}
We now ask where the molecule was at time $t$ given the entire record:$p\big(x(t)=x\mid\mathcal{D}\big)\,dx$ is the probability of finding the position in $[x,x+dx]$ at time $t$, with the photons before \emph{and} after $t$ taken into account.  Since $x(t)$ is Markovian, the present position is the only bridge between the two halves of the record: given $x(t)=x$, the photons already recorded carry no further information about those still to come.  The joint probability of position and record therefore factorizes into exactly the two filters,
\begin{equation}
  p\big(x(t)=x\mid\mathcal{D}\big)=\frac{\rho(x,t)\,\chi(x,t)}{L},
  \qquad
  L=\big\langle \rho(\cdot,t),\chi(\cdot,t)\big\rangle .
  \label{eq:smoothing_si}
\end{equation}
The normalization $L$ is the same number for every $t\in[0,T]$: it is the likelihood of the whole record, computed by splitting the trace at $t$ and integrating over the position there.  At $t=T$, where $\chi\equiv 1$, Eq.~\eqref{eq:smoothing_si} reduces to Eq.~\eqref{eq:filter}.
 
On the grid of Sec.~\ref{subsec:likelihood_eval}, $\chi$ becomes a vector $\bm\chi_n$ taken just after photon $n$, and the forward recursion of Eq.~\eqref{eq:forward} runs in reverse with transposed operators,
\begin{equation}
  \bm\chi_{N}=\mathcal{S}(\tau_{N+1})^{\top}\mathbf{1},
  \qquad
  \bm\chi_{n-1}=\mathcal{S}(\tau_{n})^{\top}\Lambda_{c_n}
  \bm\chi_{n} .
  \label{eq:backward_grid}
\end{equation}
The similarity transform of Eq.~\eqref{eq:symm} makes the two sweeps identical.  Because $A$ is symmetric, Eq.~\eqref{eq:propagator} gives $\mathcal{S}(\tau)^{\top}=\Pi^{-1/2}e^{A\tau}\,\Pi^{1/2}$.  In the transformed filters $v_n=\Pi^{-1/2}\alpha_n$ and $w_n=\Pi^{1/2}\bm\chi_n$, both sweeps therefore alternate the same two operations. First the symmetric propagator $e^{A\tau}$ and the reweighting $\Lambda_{c}$, in mirrored order. And both start from the same vector $\bm\pi^{1/2}$: the forward sweep because $\alpha_0=\bm\pi$, the backward one because $\chi(\cdot,T)=\mathbf{1}$.  In the pointwise product $\odot$ of the two filters every $\Pi^{\pm1/2}$ then cancels,
\begin{equation}
  L = w_n^{\top}v_n ,
  \qquad
  \gamma_n=\frac{w_n\odot v_n}{w_n^{\top}v_n} ,
  \label{eq:gamma}
\end{equation}
with $\gamma_n$ the posterior over the position at photon $n$ given the entire record.  Each $\gamma_n$ is a distribution over the grid cells, so posterior expectations are inner products with the grid vector $\mathbf{x}=(x_1,\dots,x_M)^{\top}$: we report the posterior mean $\hat{x}_n=\gamma_n^{\top}\mathbf{x}$.  Because $\gamma_n$ is normalized on the spot, the normalizers $Z_n$ of Eq.~\eqref{eq:forward} cancel, so both sweeps may be rescaled freely.  Evaluating $\gamma$ between the photons only requires inserting phantom events with $\Lambda=\mathbb{I}$, which pass through both recursions unchanged.

\subsection{Sampling trajectories}
Equation~\eqref{eq:gamma} is a marginal at one point in time.  A full trajectory is a draw from the joint posterior over all positions.  The Markov property factorizes this joint law backwards: given the position at photon $n{+}1$, the later photons carry no further information about the position at photon $n$.  We therefore sample exactly by forward filtering and backward sampling.  Draw the final position on the spacial grid, $i_N\sim\gamma_N$, then, for $n=N-1,\dots,1$, draw $i_n$ from the forward filter reweighted by the probability of reaching the position already drawn,
\begin{equation}
  p\big(i_n=i \mid i_{n+1}=l,\mathcal{D}\big)
  \;\propto\; [v_n]_i\big[e^{A\tau_{n+1}}\big]_{il} ,
  \label{eq:ffbs}
\end{equation}
The weight $[v_n]_i\big[e^{A\tau_{n+1}}\big]_{il}$ has a direct reading: up to the normalization over $i$, it is $[\alpha_n]_i\,[\mathcal{S}(\tau_{n+1})]_{li}$, so the forward filter times the probability of diffusing from cell $i$ to the just-drawn cell $l$ while surviving the dark gap.  The $\pi_i^{\pm1/2}$ of the two factors cancel, and the symmetry of $e^{A\tau}$ swaps the indices. The needed column of $e^{A\tau_{n+1}}$ is $\Psi\big(e^{\nu\tau_{n+1}}\!\odot\Psi_{l,:}\big)$, with $\Psi_{l,:}$ the $l$th row of $\Psi$, so a batch of paths costs two matrix--matrix products per gap and the dense propagator is never formed.  The same phantom events with $\Lambda=\mathbb{I}$ refine the sampling lattice in time, so trajectories can be drawn at any time resolution.

\clearpage
\onecolumngrid

\begin{center}
    {\Large\bfseries Supplementary Information}
\end{center}
\vspace{1em}

\setcounter{figure}{0}
\renewcommand{\thefigure}{S\arabic{figure}}

\begin{figure}[!ht]
\centering
\includegraphics[width=0.8\textwidth]{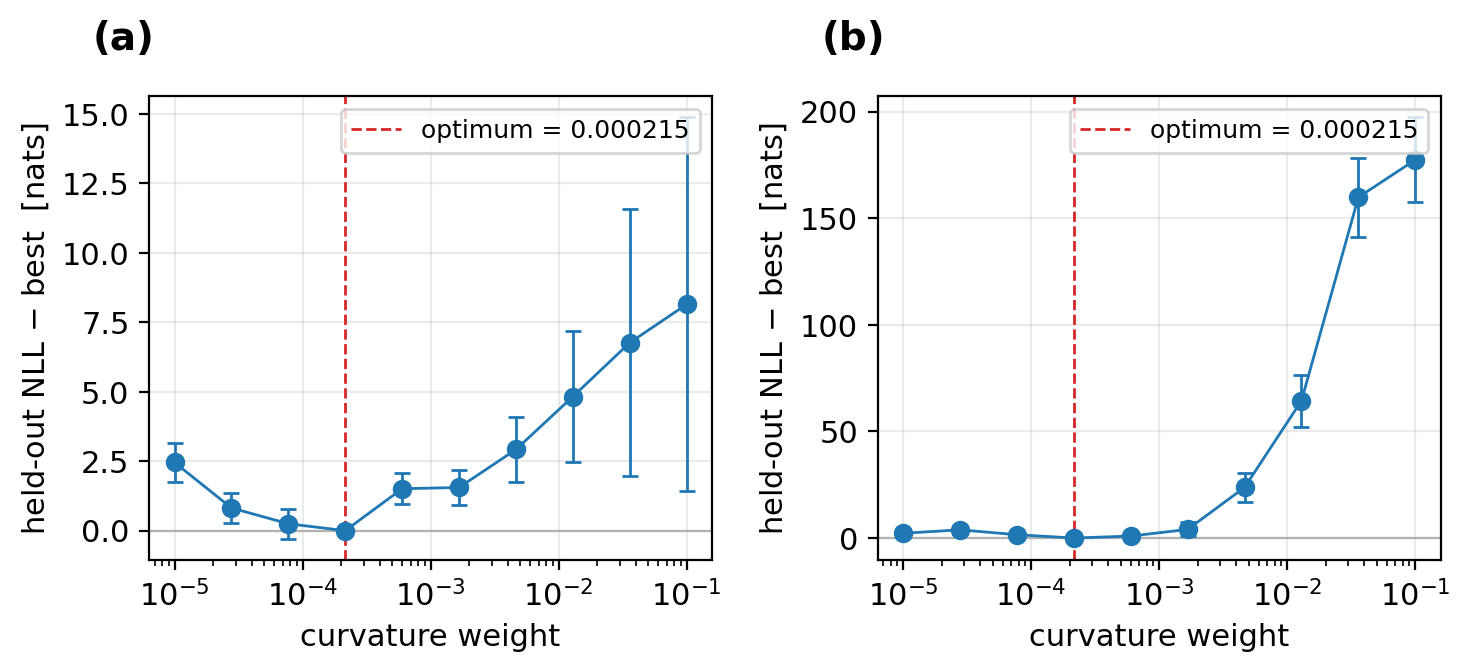}
\caption{\textbf{Curvature weight $\omega$ scan.} We performed ten optimizations on a single set of simulated trajectories, varying the curvature weight $\omega$ between runs. Each fitted model was then evaluated on five new, independent sets of simulations using the likelihood alone, i.e.\ excluding the prior. We report the sum over per-trace likelihoods, relative to the fit that achieved the lowest negative log-likelihood, together with the standard error over independent traces. The red line indicates the best value for $\omega$. (a) Shows the $\omega$ scan for the asymetric double-well, (b) shows the $\omega$ scan for the free-energy landscape with an additional intermediate state.}
\end{figure}

\begin{figure}[!ht]
\centering
\includegraphics[width=0.95\textwidth]{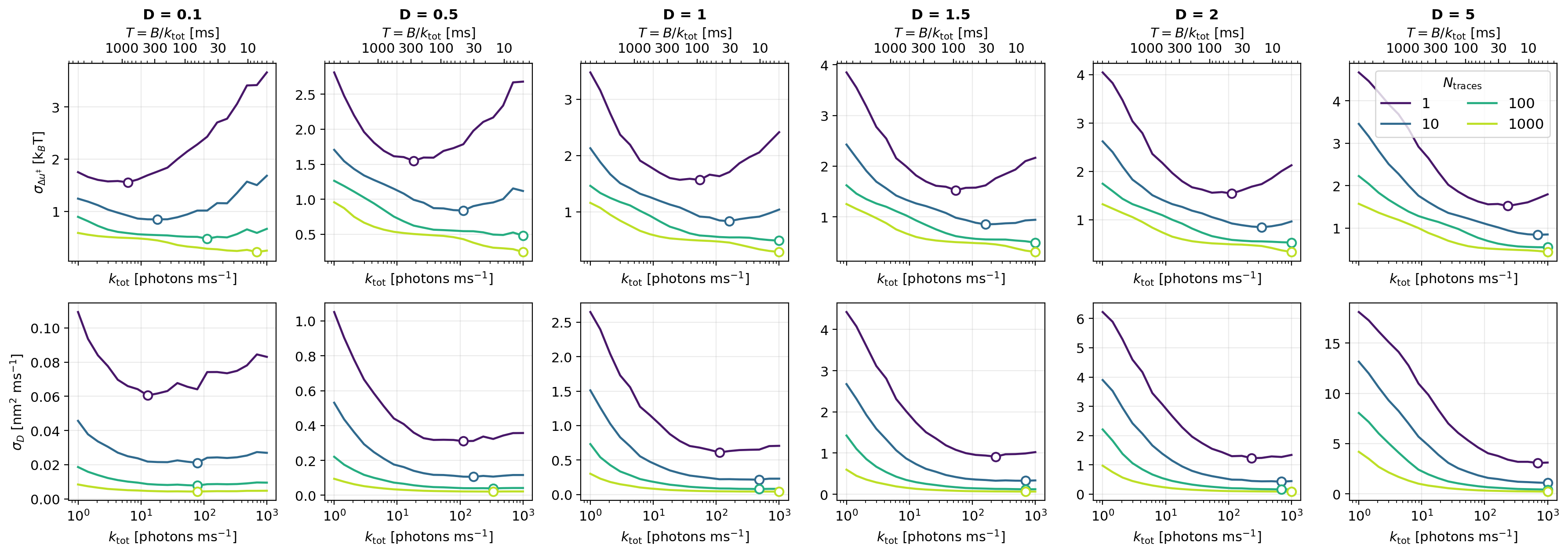}
\caption{\textbf{Optimal acquisition settings for different diffusion coefficients $D$.} We repeated the scan over $k_\mathrm{tot}$ using exactly the same settings as in the main text, changing only the underlying diffusion coefficient $D$.}
\end{figure}

\begin{figure}[!ht]
\centering
\begin{minipage}{0.55\columnwidth}
\centering
\includegraphics[width=\textwidth]{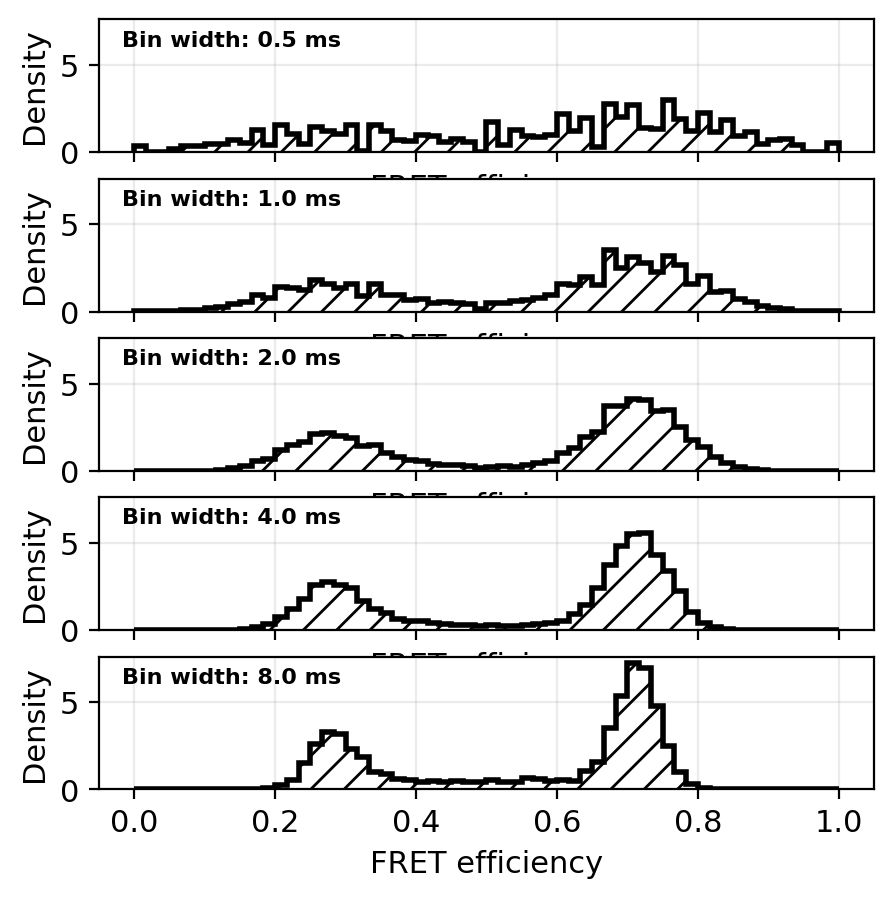}
\caption{\textbf{FRET efficiency histogram for traces with an intermediate state.} We computed the FRET efficiency histogram over all traces by dividing each trajectory into time bins and, for each bin, taking the ratio of the photons detected in the red channel to the total number of photons detected in both channels. The resulting efficiencies were then collected into a histogram. We repeated this procedure for several bin sizes.}
\end{minipage}
\end{figure}

\begin{figure}[!ht]
\centering
\begin{minipage}{0.55\columnwidth}
\centering
\includegraphics[width=\textwidth]{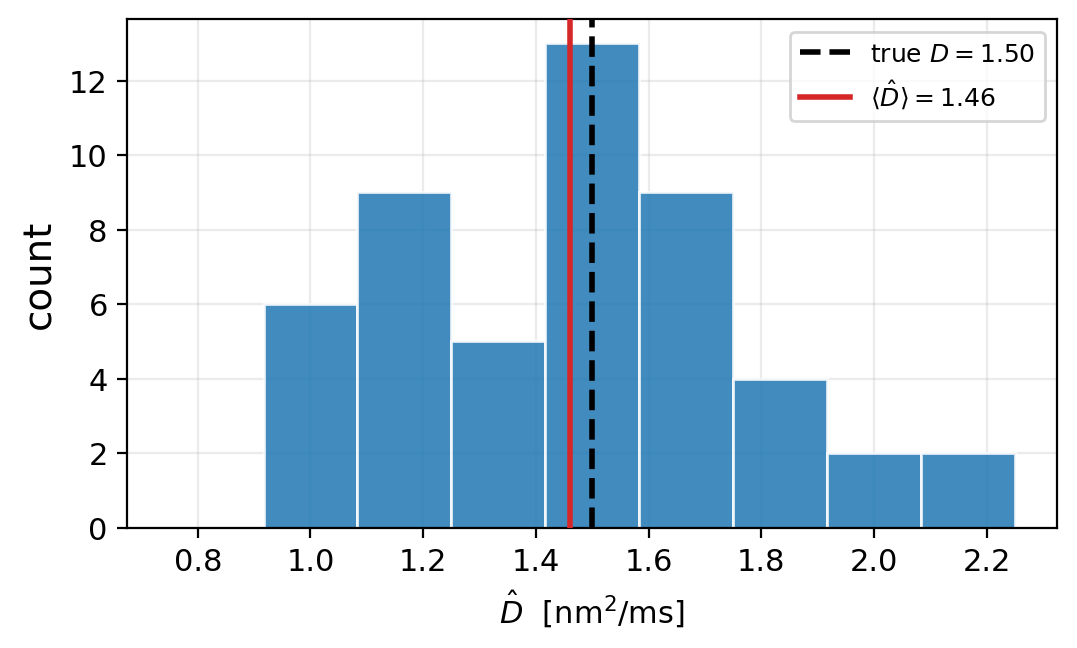}
\caption{\textbf{Diffusion coefficients obtained from many independent data sets.} Histogram of the diffusion coefficients obtained by fitting each independent data set separately. The red line marks the mean over all fitted diffusion coefficients.}
\end{minipage}
\end{figure}

\end{document}